\documentclass[slac_one]{revtex4}
\usepackage{graphicx}
\usepackage{fancyhdr}
\begin{document}

\title{Visual Physics Analysis (VISPA) - Concepts and First Applications}

\author{O.~Actis, M.~Erdmann, R.~Fischer, A.~Hinzmann, M.~Kirsch, T.~Klimkovich, G.~M\"uller, M.~Plum, J.~Steggemann\footnote{Presenter}}
\affiliation{Physikalisches Institut 3 A, RWTH Aachen University, Germany}

\begin{abstract}
VISPA is a novel development environment for high energy physics analyses, based on a combination of graphical and textual steering. The primary aim of VISPA is to support physicists in prototyping, performing, and verifying a data analysis of any complexity. We present example screenshots, and describe the underlying software concepts.
\end{abstract}

\maketitle

\thispagestyle{fancy}

\section{INTRODUCTION} 

During the past 15 years, most high energy physics experiments have been deploying object-oriented software in their experiment-specific software analysis frameworks, typically based on C++ and a specific configuration language (e.g.~\cite{Meyer:H1,Fabozzi:CMS}). Also \texttt{Python} as a dynamically-typed programming language is being used as the basis for an analysis  environment~\cite{Belyaev:LHCb}. Only recently, the issue of visualization while performing a physics analysis has become a field of strong interest. With the new project VISPA (Visual Physics Analysis), the step in between visualizing the measured objects in the detector (event display) and physics distributions (histograms) is graphically supported. In this step, the primary tasks of a physicist are prototyping, executing, and verifying a physics analysis. This is an iterative procedure until the analysis is finalized.

VISPA facilitates prototyping, performing, and verifying a physics data analysis by combining graphical and textual programming. This combination has been shown to speed up design and development in other fields, e.g. hardware control using the \texttt{LabView} program \cite{Labview}. To deploy a visual environment for physics analysis, VISPA provides a multi-purpose window tool with a three column structure, a navigator panel, a window for graphical displays, and a property panel. For the text-based programming, both the C++ and the \texttt{Python} languages are supported.

VISPA has been developed independently of experiment-specific software with a well defined interface to connect to all high energy physics experiments. The interface is based on the software package \texttt{PXL} which provides a general container to hold all relevant information of a high energy physics event  \cite{bib_pxl}.

\begin{figure}[ht]
\centering
\includegraphics[width=160mm]{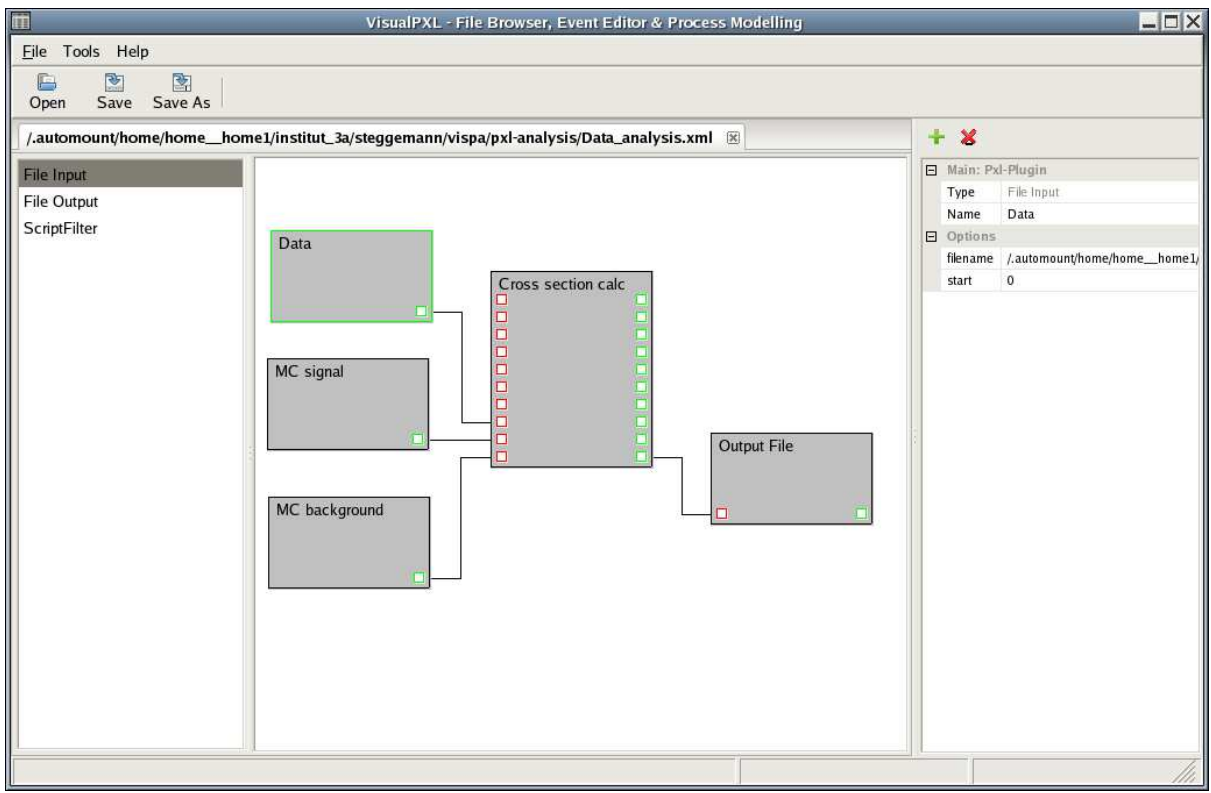}
\caption{Analysis design with the VISPA graphical platform. \label{browser_design}}
\end{figure}

For designing a physics analysis, VISPA provides a graphical module steering which enables physicists to add, connect, and configure the analysis modules. Based on a plug-in mechanism, modules are already provided for the purpose of reading/writing event data, or for individual user analysis code. The latter can be coded directly within a corresponding editor. The visualization of the module steering facilitates verification of the analysis structure and communication between physicists doing analysis (Fig.~\ref{browser_design}). When using the \texttt{Python} language in their analysis code, the user programs can be executed without compiling and linking.

\begin{figure}[ht]
\centering 
\includegraphics[width=160mm]{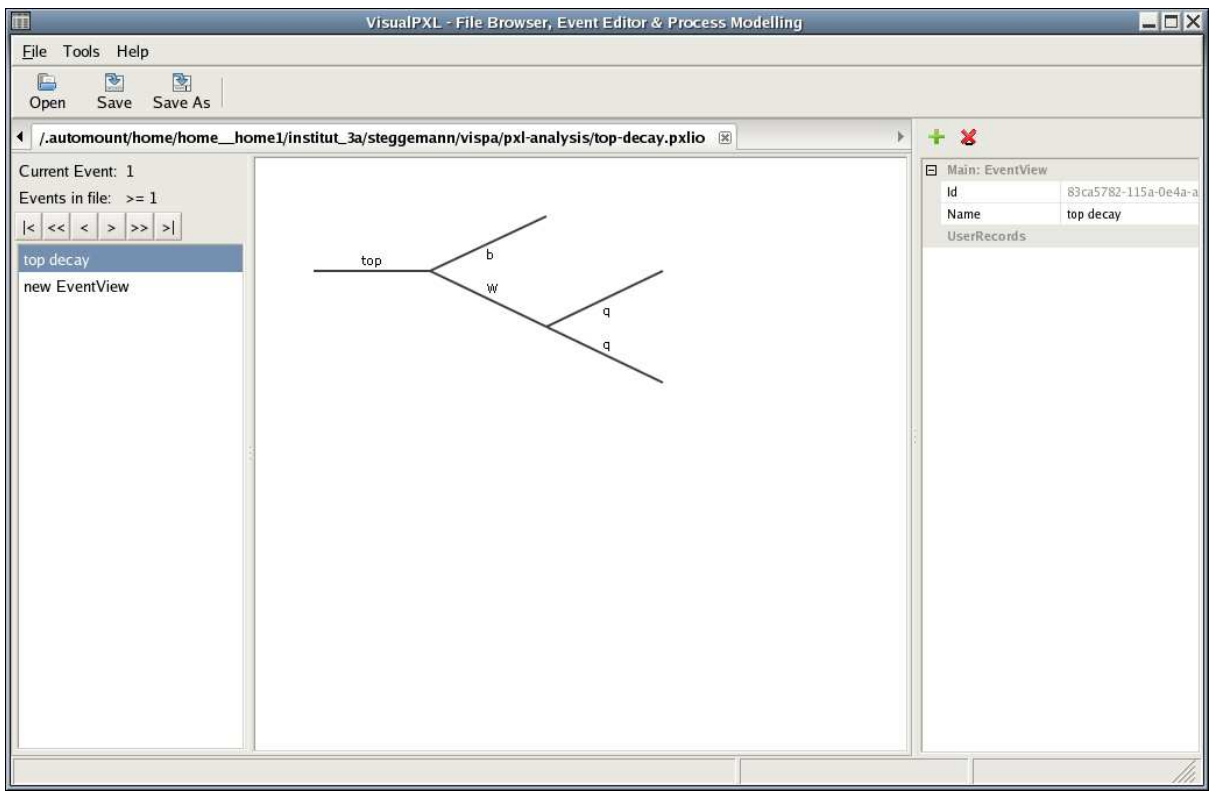}
\caption{Visualization of a particle decay tree in the VISPA event browser. \label{browser_decay}}
\end{figure}

For verification of the analysis, VISPA provides an event browser to inspect the overall event structure, to display particle decay cascades, and the properties of each particle (Fig.~\ref{browser_decay}).

This contribution is organized as follows: first, the underlying C++ toolkit \texttt{PXL} is described. Then, the module steering system is discussed, followed by a description of the available analysis modules. Finally, the VISPA graphical user interface is detailed.

\section{THE PXL TOOLKIT}

The C++ toolkit \texttt{PXL} has been developed since 2006 \cite{bib_pxl}. It is the successor of the PAX toolkit, which was developed from 2002 to 2007 \cite{Erdmann:2002wn, Kappler:2005tf}. \texttt{PXL} provides all necessary features for an experiment-independent high level physics analysis with emphasis on an easy user syntax. Particularly, \texttt{PXL} enables the reconstruction of decay trees and the handling of analyses with reconstruction ambiguities.

\texttt{PXL} offers an extensible collection of physics objects, representing particles, vertices, events, and collisions. In the analysis of an event containing reconstructed data, new information can be added to each object by means of user event data. Between all objects, relations can be established, e.g. to build up decay trees, or to associate reconstructed with generated particles. The classes \texttt{pxl::Event} and \texttt{pxl::EventView} represent a whole physics event and a special view thereof, respectively. They act as containers for physics objects, holding the relations between them, and as the standard interface to algorithms. Copies of these classes preserve all contained information such as the relations between particles. These features efficiently support the evaluation of hypotheses. \texttt{PXL} includes all objects into its I/O scheme, where a file is made up of compressed binary data chunks, thus guaranteeing file integrity during job execution.

In addition to the core features, we also offer a \texttt{Python} extension, \texttt{PyPXL}, to enable the usage of all \texttt{PXL} objects and their methods within \texttt{Python} programs.

\section{MODULE STEERING}

The design of physics analyses with VISPA is based on the decomposition of analyses into modules. Whereas a simple analysis may require only a few modules which are connected serially, a complex analysis requires more modules and more sophisticated streams of data.

The VISPA module steering system controls the data flow as well as the selection and settings of analysis modules. The module selection is based on a plug-in mechanism, guaranteeing extensibility and efficiency as only libraries for the analysis modules deployed in the current analysis need to be provided. The data flow is managed by connections between sources (data input) and sinks (data output) of the respective modules. This permits the usage of multiple data streams as needed in complex physics analyses. Any state of the analysis can be stored and received back either in \texttt{XML} format or in \texttt{Python}. The analysis can be executed either in batch mode, or interactively from the VISPA graphical platform.

\section{ANALYSIS MODULES}

Analysis modules for tasks of different complexity within a physics analysis exist. File operations are handled by input and output file modules. For user-demanded tasks within an analysis, a generic \texttt{Python} analysis module is provided, where the \texttt{Python} analysis code can be edited interactively within the VISPA graphical platform. 

An example of a complex analysis module is the automated reconstruction of particle cascades, a task arising in the reconstruction of particles with combinatorial ambiguities \cite{Actis:2008tf}. Given a template of a particle decay cascade and reconstructed particle data as input, this module generates all possible reconstruction versions. For events simulated with a Monte Carlo generator it supports finding the reconstructed version corresponding to the correct decay chain.

\section{GRAPHICAL USER INTERFACE}

The VISPA multi-purpose window tool serves as the graphical user interface for the design and steering of analyses, and for the browsing of complete physics events. It provides a common user interface for these different tasks, see Figs.~\ref{browser_design} and~\ref{browser_decay}. The frame on the left-hand side is used for the selection of modules or events, the main window displays the analysis modules or the event content, and the frame on the right-hand side shows the properties of the item selected in the main window. All objects in the main frame can be selected and moved like in popular software (drag-and-drop). Icons for opening and saving data files as well as other analysis modules are provided. By using a tabbed document interface, several files can be opened in parallel.

Browsing physics event data allows the verification of physics analyses on an event-by-event basis. The correctness of each object can be explored by selecting it, and inspecting its contents in the property grid on the right-hand side. In addition, the visualization of decay trees allows to check if all relations have been established correctly.

To display complex decay trees, the VISPA graphical browser incorporates an algorithm for their proper layout. The algorithm is based on a model of physical forces, like spring forces, or gravity. Each starting and end point of a particle is provided with a node subjected to these forces. Using an iterative procedure, the positions of the nodes are optimized with respect to balanced forces. This algorithm results in a well-distributed view of, e.g., asymmetric decay trees.

\section{CONCLUSION}

A novel physics analysis environment, VISPA, has been presented. VISPA facilitates prototyping, performing, and verifying a physics data analysis by combining graphical and textual programming. It provides the so far missing graphical support for physicists in the step between displaying events, and visualizing physics distributions.

\begin{acknowledgments}

We are very grateful for financial support of the Ministerium f\"{u}r Innovation, Wissenschaft, Forschung und Technologie des Landes
Nordrhein-Westfalen, the Bundesministerium f\"{u}r Bildung und Forschung (BMBF), and the Deutsche Forschungsgemeinschaft (DFG).

\end{acknowledgments}

\end{document}